\PassOptionsToPackage{hyphens}{url}
\documentclass{river-journal}
\usepackage{rivps}
\usepackage[hidelinks]{hyperref}
\usepackage{latexsym, amssymb, amsmath, amsfonts, amsthm, bm, bbm, nicefrac, stmaryrd, thmtools, enumitem, centernot}
\usepackage{url, graphicx, xcolor, rotating}
\usepackage{appendix}

\definecolor{svlinks}{rgb}{.0,0.3,0.6}
\usepackage[hang]{footmisc}
\usepackage[T2A,T1]{fontenc}
\usepackage[utf8]{inputenc}
\usepackage[russian,greek,UKenglish]{babel}

\usepackage{epsfig, listings, appendix}
\usepackage{multirow}
\usepackage{float}
\usepackage{placeins}
\usepackage[all]{nowidow}
\usepackage{fnpct}

\usepackage{tikz, wrapfig, caption}
\usetikzlibrary{matrix}
\usepackage[normalem]{ulem}
\usepackage{booktabs}
\usepackage{array}
\usepackage[noshell]{gnuplottex}

\setlist{nosep}

\usepackage{etoolbox}
\pretocmd{\eqref}{Eq.~}{}{}

\par

\begin{document}
\begin{opening}
\title[Typosquatting for Fun and Profit]{Typosquatting for Fun and Profit: Cross-Country~Analysis of Pop-Up Scam}

\author{Tobias Dam\textsuperscript{1,3}, Lukas Daniel Klausner\textsuperscript{1,4}, Sebastian Schrittwieser\textsuperscript{2,5}}
\institute{
\textsuperscript{1}{
Institute of IT Security Research\\
St.\ P\"olten University of Applied Sciences\\
Austria\\}
\textsuperscript{2}{
Josef Ressel Center TARGET\\
St.\ P\"olten University of Applied Sciences\\
Austria\\
}
\textsuperscript{3}{tobias.dam@fhstp.ac.at}\\
\textsuperscript{4}{mail@l17r.eu}\\
\textsuperscript{5}{sebastian.schrittwieser@fhstp.ac.at}
}

\end{opening}

\runningtitle{Typosquatting for Fun and Profit}
\runningauthor{Dam, Klausner, Schrittwieser}


\subsection*{Abstract}
Today, many different types of scams can be found on the internet. Online criminals are always finding new creative ways to trick internet users, be it in the form of lottery scams, downloading scam apps for smartphones or fake gambling websites. This paper presents a large-scale study on one particular delivery method of online scam: pop-up scam on typosquatting domains. Typosquatting describes the concept of registering domains which are very similar to existing ones while deliberately containing common typing errors; these domains are then used to trick online users while under the belief of browsing the intended website. Pop-up scam uses JavaScript alert boxes to present a message which attracts the user's attention very effectively, as they are a blocking user interface element.

Our study among typosquatting domains derived from the Majestic Million list utilising an Austrian IP address revealed on 1\,219 distinct typosquatting URLs a total of 2\,577 pop-up messages, out of which 1\,538 were malicious. Approximately a third of those distinct URLs (403) were targeted and displayed pop-up messages to one specific HTTP user agent only. Based on our scans, we present an in-depth analysis as well as a detailed classification of different targeting parameters (user agent and language) which triggered varying kinds of pop-up scams. Furthermore, we expound the differences of current pop-up scam characteristics in comparison with a previous scan performed in late 2018 and examine the use of IDN homograph attacks as well as the application of message localisation using additional scans with IP addresses from the United States and Japan.



\section{Introduction}
\label{sec:introduction}

Pop-up ads have been an annoying phenomenon on the internet since the 1990s. This type of web advertisement puts the ad banner into a separate browser window instead of directly integrating it with the website~\cite{rodgers2000interactive}. The great popularity of pop-up ads among advertisement companies contrasted with the dissatisfaction of users because of the ads' highly intrusive nature. This eventually caused all major browser vendors to implement pop-up blockers in their software in the early 2000s~\cite{hansell_2004}. Today, websites rarely make use of pop-ups and pop-up ads have disappeared almost completely from the web (as browsers would block them anyway).

However, similar concepts are now being used for online scams. Instead of displaying an ad or malicious content in a new browser window through the JavaScript method \texttt{window.open}\footnote{\url{https://developer.mozilla.org/en-US/docs/Web/API/Window/open} (last accessed: 30 November 2019)}, a new trend in web-based scams can be observed: The JavaScript method \texttt{alert}\footnote{\url{https://developer.mozilla.org/en-US/docs/Web/API/Window/alert} (last accessed: 30 November 2019)} is used to show a short text message to the user.

Displaying the phishing message inside a JavaScript alert box has one important advantage for the attacker: An alert box steals the focus of the entire website. While normal advertisements can easily be ignored, alert boxes require the user to actively click a button to dismiss them. This obligatory interaction combined with the often short messages creates an effective entry point to further engage the user. This initial forced attention can then be exploited to lure the user to a dedicated website which serves the attacker's purpose, e.\,g.\ by asking for email addresses or credit card details. Attackers have also been observed repeatedly opening alert boxes, trying to pose as legitimate OS error messages and scaring the user into thinking that their device has been infected by malware~\cite{miramirkhani2016dial}.

These properties make alert boxes a very effective and widely abused vector for attackers. However, little attention has been paid to the described techniques by the research community. We conducted a scan utilising typosquatting URLs based on the Alexa Top 1~Million websites~\cite{alexa} in late 2018 and performed a large-scale analysis~\cite{popup_scam} using the collected data. During our work, we discovered various hints of localisation attempts, which were likely caused by our Austrian IP address. In order to examine whether our assumption was correct as well as to observe possible changes of pop-up message characteristics since our last scan, we decided to perform additional scans utilising IP addresses from Austria, the United States and Japan.

Based on the Majestic Million websites~\cite{majestic}, we created a list of websites with commonly misspelt names. This set consisted of 638\,835 valid, registered domain names, which we scanned using automated browsers with five different user agents. In this paper, we present an up-to-date comprehensive, large-scale study of the use of automatically displayed pop-up scams on websites and analyse how different user agents and languages are targeted by these campaigns. Additionally, we compare our results to our prior pop-up scam analysis research.

In particular, the main contributions of this paper are:
\begin{itemize}
    \item We present an up-to-date comprehensive scientific large-scale follow-up study of the utilisation of JavaScript pop-up messages for online scams on typosquatting URLs based on the Majestic Million websites.
    \item In contrast to our previous scan, we also include IDN homograph attacks as a possible typosquatting attack vector.
    \item We provide insight into the goals and purposes of the pop-up messages and the sites hosting them by manually defining and assigning categories based on the message content and the websites.
    \item Various distributions of the languages and the user agents across the different distinct messages, websites and categories are visualised and detailed, in order to explain the current state as well as trends in this particular delivery method for online scams.
    \item A comparison with our previous scan in relation to pop-up scams details the developments and the differences of the messages as well as the sites displaying them.
    \item An assessment of the application of alert box message localisation by utilising IP addresses from three different countries.
\end{itemize}

The remainder of this paper is structured as follows: We discuss related work in \autoref{sec:relatedwork} and give a technical overview of the utilised framework in \autoref{sec:technical}. We present the results of our research in \autoref{sec:results} and evaluate the scan results in \autoref{sec:evaluation}, where we also present a large-scale analysis. Possible future work is detailed in \autoref{future work} and \autoref{sec:conclusion} concludes the paper. All diagrams visualising our results are located in the appendix.
\section{Related Work}
\label{sec:relatedwork}

One important online scam category is phishing. It has been around for a long time as one of the most effective social engineering techniques and is a well-studied research area (see e.\,g.\ \cite{abraham2010overview,mohammad2013predicting,mohammad2014predicting}). Due to the fact that the majority of today's users have only limited technical and security knowledge, the success rate of social engineering attacks is constantly high. Moreover, adversaries are becoming more and more creative in handcrafting their attacks to increase their success rate. While traditional means of delivery (i.\,e.\ via email~\cite{NLPPhishing}) are still widely used, many other delivery methods exist. Typosquatting~\cite{typosquat2003} (also referred to as ``URL hijacking'') is a technique which is based on the concept of registering domain names with typing errors and similar mistakes made by users when entering a popular web address.

One of the first large-scale studies on typosquatting was conducted in 2003 by Edelman~\cite{Edelman}, who discovered more than 8\,800 registered domains which were typographical variations of the most popular domain names at that time. His findings showed that most of those domain names were traced back to one individual, John Zuccarini, who used these typosquatted domains to redirect users to websites containing sexually explicit content. Furthermore, he was found to use particular tactics to trap the users from leaving these sites, such as blocking the browser's ``Back'' and ``Close'' functionalities. 

Typosquatting attacks are based on the insertion, deletion or substitution of characters or the permutation of adjacent characters in popular domain names~\cite{kintis2017hiding}. Holgers et al.~\cite{Holgers:2006} conducted an experiment in 2006 in which they measured the effect of visual similarities between letters in particular domain names. At that time, their results outlined that such homograph attacks were very rare and not severe in nature. However, the increasing use of internationalised domain names (IDNs) as well as the rising number of malicious IDN registrations over the last years show the increasing significance of this typosquatting technique~\cite{reexamination-idn, farsight}. 

Numerous other squatting techniques such as \emph{bitsquatting}, \emph{combosquatting}, and \emph{soundsquatting} were thoroughly researched in the past. Bitsquatting is the act of registering a domain name one bit different than an original domain, which might be accessed by users due to bit errors changing their memory content. Dinaburg~\cite{dinaburg} performed an experiment in which he registered 30 bitsquatted versions of popular domains and logged all HTTP requests. His findings outlined that there were 52\,317 bitsquat requests from 12\,949 unique IP addresses over the course of eight months. Nikiforakis et al.~\cite{Nikiforakis:2013} conducted one of the first large-scale analyses of the bitsquatting phenomenon. Their results clearly showed that new bitsquatting domains are registered daily and are commonly used by the adversaries for generating profit through the use of ads, abuse of affiliate programs and, in some cases, distribution of malicious content.

Kintis et al.~\cite{kintis2017hiding} conducted a study on combosquatting, which combines brand names with other keywords in the domain names. Their study showed that combosquatting domains are widely used to perform various types of attacks, including phishing, social engineering, affiliate abuse, trademark abuse and malware.

Furthermore, Nikiforakis et al.~\cite{soundsquatting} presented a concept called soundsquatting which takes advantage of user confusion over homophones and near-homophones, i.~e.\ words which sound similar or the same, but are spelled differently. To verify how much this soundsquatting technique is used in the wild, Nikiforakis et al.\ developed a tool to generate possible soundsquatted domains from a list of target domains. Using the Alexa Top~10,000 websites, they were able to generate 8\,476 soundsquatted domains out of which 1\,823 were already registered. 

Additionally, Nikiforakis et al.~\cite{Nikiforakis:2012} conducted a study in which they examined malicious JavaScript inclusions. Their findings included a vulnerability which occurs when a developer mistypes the address of a JavaScript library in their HTML pages. This would allow an attacker to easily register the typosquatted domain which could then compromise the website including a malicious JavaScript library.

Pop-up scam has not been researched in much detail yet. Miramirkhani et al.~\cite{miramirkhani2016dial} performed a large-scale analysis of one particular type of pop-up scams, namely technical support scams. Their methodology included a check for Java\-Script alert boxes. In Chou et al.'s work~\cite{chou2008detecting} the detection of traditional (JavaScript-less) pop-up ads through machine learning was proposed. The psychological aspects of fake pop-ups on internet users were analysed by Sharek et al.~\cite{sharek2008failure}.

We performed a large-scale analysis of pop-up scams on typosquatting URLs~\cite{popup_scam} by utilising a modified version of the 
\textsc{MiningHunter} 
framework and an Austrian IP address. Our scan on the typosquatting domains based on the Alexa Top 1~Million list revealed on 8\,255 distinct typosquatting URLs a total of 9\,857 pop-up messages, out of which 8\,828 were malicious. We found that the majority of distinct URLs were targeting mobile browser user agents. Additionally, we categorised the messages and provided insights on their characteristics and goals as well as the application of localisation.

\section{Technical Overview}
\label{sec:technical}

To perform the large-scale scans required for this research, we employed a modified version of the
\textsc{MiningHunter} \cite{mininghunter}
framework, which we initially developed to
identify browser-based cryptocurrency mining campaigns.
\textsc{MiningHunter}
is based on Docker Swarm\footnote{\url{https://docs.docker.com/engine/swarm/key-concepts} (last accessed: 30 November 2019)} and consists of automated browsers and a back end where the collected data is stored.

To scan websites at a large scale, a Chromium browser installed inside a Docker container is automated using the Chrome DevTools protocol\footnote{\url{https://chromedevtools.github.io/devtools-protocol} (last accessed: 30 November 2019)}. It receives scanning requests via a Kue\footnote{\url{https://github.com/Automattic/kue} (last accessed: 30 November 2019)} job queue, automatically loads the website and records various details such as visited URLs. The accumulated data is then sent to a back end container through HTTPS and stored inside a MongoDB\footnote{\url{https://www.mongodb.com/what-is-mongodb} (last accessed: 30 November 2019)} database for later analysis. To scan a large number of websites within a reasonable time span, multiple scanning containers can be active at the same time.

For the purpose of testing we mimicked the most common behaviour of an adversary, namely, we made use of a technique popularly known as ``typosquatting'', as explained in \autoref{sec:relatedwork}. In our experiment, we applied this technique to the Majestic Million websites. To be able to cover the broad spectrum of the web address permutations, we used \texttt{dnsmorph},\footnote{\url{https://github.com/netevert/dnsmorph} (last accessed: 30 November 2019)} a tool which generates possible typosquatting URLs for a particular URL. From the pool of thousands of possible address permutations, we selected only those which were actually registered as valid domains (in total, we were able to generate and verify 638\,835 registered domain names). In contrast to \texttt{dnstwist},\footnote{\url{https://github.com/elceef/dnstwist} (last accessed: 30 November 2019)} which was used in our first study, \texttt{dnsmorph} is able to generate internationalised domain names (IDN) homograph attacks (using homographs such as Greek omicron `\foreignlanguage{greek}{ο}' or Cyrillic es `\foreignlanguage{russian}{с}'). However, this attack vector seems to be purely theoretical in our use case, as we could not find any application of that attack using our scans -- the generated list of registered typosquatting domains did not contain any IDN homographs, at all.

For the purpose of this research, we developed two additional custom plugins for our framework. The first plugin, \textit{UserAgentSpoofer}, sends a configurable, fake user agent to allow us to discern differences in behaviour which depend on this HTTP header. The plugin replaces the \texttt{User-Agent} request header in all requests sent to websites using the \texttt{Network.setUserAgentOverride} method of the Chrome DevTools protocol. The second plugin, \textit{AlertRecorder}, stores URLs and messages of all JavaScript alert boxes encountered while loading and rendering a website. The data is acquired using the \texttt{Page.javascriptDialogOpening} API.

Websites are scanned until the \texttt{Network.loadingFinished} event is triggered by the Chrome DevTools protocol, plus an additional second in order to capture alerts that appear after the site has finished loading. The scan is also stopped in case the \texttt{Network.loadingFinished} event is not triggered 30~seconds after beginning to load the website.

Using these two plugins, we performed five full scans of our list of typosquatting domains based on the Majestic Million websites. To be able to provide a wider variety of targets, each scan used a different user agent. We updated the user agents used in our first scan according to current browsers and operating systems and selected Chrome~78 and Firefox~70 (both from 2019) to represent two popular, modern browsers running on Windows~10. We additionally included Internet Explorer~11 (from 2015) on Windows~10 to determine if any campaigns specifically target Microsoft's default browser for that OS. To cover the most commonly used mobile devices, we included Chrome~77 on Android~9 and Safari~13 on iOS~13.1 (both from 2019). Detailed information regarding all user agents selected for the scans can be found in \autoref{table:useragents}.

For the purpose of analysing localisation attempts, the scans of the five different user agents were repeated with IP addresses from other countries. The traffic of the automated browser inside the Docker containers was routed through an OpenVPN\footnote{\url{https://openvpn.net/} (last accessed: 30 November 2019)} tunnel to endpoints in specific countries provided by AirVPN\footnote{\url{https://airvpn.org/} (last accessed: 30 November 2019)}.

\begin{sidewaystable}
    \centering
    \begin{tabular}{|c|m{9cm}|c|c|}
    \hline
    \textbf{label} & \multicolumn{1}{|c|}{\textbf{user agent}} & \textbf{operating system} & \textbf{browser} \\
    \hline
    chrome & Mozilla/5.0 (Windows NT 10.0; Win64; x64) AppleWebKit/537.36 (KHTML, like Gecko) Chrome/78.0.3904.70 Safari/537.36 & Windows 10 & Chrome 78 \\
    \hline
    ie & Mozilla/5.0 (Windows NT 10.0; WOW64; Trident/7.0; rv:11.0) like Gecko & Windows 10 & Internet Explorer 11 \\
    \hline
    iossafari & Mozilla/5.0 (iPhone; CPU iPhone OS 13\_1\_3 like Mac OS X) AppleWebKit/605.1.15 (KHTML, like Gecko) Version/13.0.1 Mobile/15E148 Safari/604.1 & iOS 13.1 & Safari 13 \\
    \hline
    firefox & Mozilla/5.0 (Windows NT 10.0; Win64; x64; rv:70.0) Gecko/20100101 Firefox/70.0 & Windows 10 & Firefox 70 \\
    \hline
    androidchrome & Mozilla/5.0 (Linux; Android 9; TA-1053) AppleWebKit/537.36 (KHTML, like Gecko) Chrome/77.0.3865.116 Mobile Safari/537.36 & Android 9 & Chrome 77 \\
    \hline
    \end{tabular}
    \caption{The user agents used for the scans. ``Label'' is a unique identifier used throughout this paper when referring to the corresponding user agent. ``Operating system'' and ``browser'' refer to corresponding technology implied by the user agent.}
    \label{table:useragents}
\end{sidewaystable}

\section{Results}
\label{sec:results}
As mentioned in \autoref{sec:introduction}, our previous scan revealed several signs of message localisation. In order to assess whether this result was caused by our Austrian IP address, we performed additional scans utilising IP addresses of two further countries via a VPN service. 

Our scans originating from Austrian IP addresses (utilising different user agents as described in \autoref{sec:technical}) resulted in a total of $2\,577$ recorded alert boxes as well as $1\,219$ distinct URLs and $303$ distinct messages. $1\,538$ of the recorded alert boxes can be considered malicious. An interesting aspect of our results is the targeting of specific user agents, which is further detailed in \autoref{sec:evaluation}: $403$ websites displayed an alert box only to one particular user agent, whereas $816$ websites showed messages to more than one user agent. Considering only distinct messages, we observed similar behaviour -- $102$ distinct messages were only shown to one particular user agent, $201$ to more than one.

The second scan used IP addresses from the United States and yielded $2\,420$ recorded alert boxes. It consisted of $1\,155$ distinct sites, $298$ distinct messages and $1\,393$ potentially malicious alert boxes. The results regarding the user agent aspect are quite similiar to the Austrian scan: $382$ websites showed an alert box only to one particular user agent, while $773$ websites served them to multiple user agents. $94$ distinct messages were used for only one specific user agent, $204$ for more than one.

The third scan used IP addresses from Japan, during which we recorded $2\,403$ alert boxes, $305$ distinct messages and $1\,098$ distinct sites. $1\,403$ alert boxes can be considered malicious. The results of this scan feature similar characteristics regarding the user agent as the aforementioned scans: $336$ websites displayed an alert box only to one particular user agent, while $762$ websites showed messages to more than one user agent. $103$ distinct messages were only shown to one particular user agent, $202$ to more than one.

The aforementioned statistics show that the number of sites displaying alert boxes declined for each scan, although each scan used the same set of typoquatting URLs. Due to limited resources, we performed the Austrian, US and Japanese scans consecutively over four weeks. We found that several domains are registered only for a short period of time, and deregistered or used for other purposes afterwards. This might be caused by e.\,g.\ phishing site takedowns or attempts to remain anonymous by the website operators.
\section{Evaluation}
\label{sec:evaluation}

Using the categories described in \autoref{sub:categories} as well as the user agents shown in \autoref{table:useragents}, we determined specific characteristics of the recorded alert box messages with respect to these features.

\subsection{Categories}
\label{sub:categories}

In order to determine which websites try to achieve similar goals by displaying a message inside an alert box as well as to enable clearer visual representations of the distribution of message types across different user agents, we selected a number of categories from our findings and assigned one to each message.

Most messages in the \textsc{Fraud} category declare that the user's device is infected by a virus or inform you about a pending update. They either urge the user to download potentially malicious software or to call a specified phone number. These characteristics contrast with those of the \textsc{Fraud} messages in our previous study, where most messages claimed that the user would receive some free credit for gambling if they registered and provided their credit card information. 

Messages contained in the \textsc{Lottery} category either claim the visitor has already won a lottery or that they have a particularly high chance of winning one. Such websites often either require the user to play a ``game'', such as spinning a wheel of fortune, or to answer questions regarding the prize (e.\,g.\ a smartphone). After completing such tasks, the websites reveal that the prize is actually a ``special offer'' and ask the visitor to provide their credit card information.

All messages in the category \textsc{APK} are in Chinese and most of them urge the user to download a dedicated application for displaying adult content. Unlike alert boxes in the category \textsc{Mobile Client}, they do not redirect to app store websites, but instead offer a direct download of an Android APK file or redirect to an iOS \texttt{itms-services} URL. Several samples were analysed using VirusTotal\footnote{\url{https://www.virustotal.com} (last accessed: 30 November 2019)} and were identified as potentially unwanted programs (such as adware and spyware) as well as Trojans.

Based on the characteristics of the alert box message content as well as manual inspection of selected samples for each distinct message, we consider messages inside the categories \textsc{Fraud}, \textsc{Lottery}, and \textsc{APK} to be malicious (e.\,g.\ phishing). Besides these malicious categories, we further defined various non-malicious categories; they were differentiated by content and message purpose in order to gain additional insight into the reasons for showing alert boxes in general.

The category \textsc{Errors} contains several types of error messages, e.\,g.\ indicating invalid access tokens or JavaScript errors as well as website maintenance and discontinuation notices.

Messages categorised as \textsc{Download} urge the user to install or update either Java or Adobe Flash Player and redirect the user to the corresponding download area. Manual inspection of the websites included in our scan which displayed these messages showed that the alert boxes do actually redirect to the legitimate websites of the software manufacturers.

\textsc{Adult} messages inform the user about adult content on the visited website, ask the user to confirm that they are of legal age and present the website's terms and conditions.

Messages of the category \textsc{Mobile Site} ask the visitor whether they want to display the dedicated mobile version of the website.

\textsc{Mobile Client} messages inform the user about the website's smartphone app and redirect the user to the according app store website.

The categories \textsc{Mobile Site} and \textsc{Mobile Client} are combined into the category \textsc{Mobile} in diagrams throughout this paper.

Messages of the \textsc{Gambling} category are related to gambling websites. All messages are in Chinese, and most websites hosting these alert boxes provide the latest results of the Hong Kong Jockey Club's Mark Six lottery\footnote{\url{https://bet.hkjc.com/marksix/} (last accessed: 30 November 2019)} as well as other gambling information. They either require the user to register on a different website or present a special offer along with an ID or contact number for instant messengers, which are in widespread use in the People's Republic of China. Since the websites did not directly request credit card information or deceive the visitors in other ways, and since we could not easily investigate the associated instant messenger accounts, we chose to separate these messages into the category \textsc{Gambling} instead of including them in the more explicitly malicious category \textsc{Lottery}.

\textsc{Misc} categorises alert box contents which do not fit into any other category and include short cookie policy statements, welcome messages and password prompts as well as various other kinds of miscellaneous messages.

\subsection{Analysis}
\label{sub:analysis}

Our results show that a significant portion of the scanned websites target visitors with mobile web browser user agents. As Figures~\ref{fig:uniquemessages}, \ref{fig:uniquemessages_us} and \ref{fig:uniquemessages_jp}, which are located in the appendix, illustrate, there are few distinct messages displayed only to a specific user agent. While the difference in the number of messages only shown to one specific user agent is not significant, there is a large disparity between the number of websites focussing on desktop web browsers and those targeting mobile web browsers in general.

\begin{figure}[hbt!]
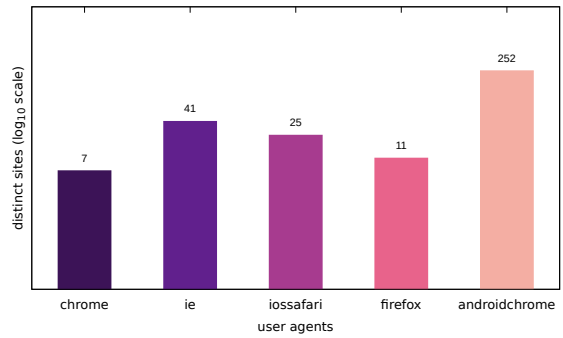

    \centering
    \begin{gnuplot}[terminal=pdf,scale=0.6]
            set boxwidth 0.5
            set style fill solid
            set yrange [0.1:2500]
            set logscale y
            unset ytics
            unset key
            set xlabel "user agents"
            set ylabel "distinct sites (log_{10} scale)"
            set linetype 1 lc rgb '#3c1357'
            set linetype 2 lc rgb '#61208d'
            set linetype 3 lc rgb '#a73b8f'
            set linetype 4 lc rgb '#e8638b'
            set linetype 5 lc rgb '#f4aea3'
            plot "Data_jp/unique_urls_per_useragent" using 1:3:($0+1):xtic(2) title "unique urls per useragent" with boxes lc variable,"Data_jp/unique_urls_per_useragent" using 1:($3+$3*0.6):3 with labels font "Arial,10"
        \end{gnuplot}
        \caption{Number of websites targeting one \\ particular user agent (Japanese IP address)}
        \label{fig:uniquesites_jp}
\end{figure}

Figures~\ref{fig:uniquesites}, \ref{fig:uniquesites_us} and \ref{fig:uniquesites_jp} illustrate the number of websites which displayed an alert box only to one particular user agent. While some alert boxes are legitimately directed at specific user groups, the number of alert boxes shown only to desktop browsers is almost negligible. We present some possible explanations for the specific targeting of mobile users in \autoref{sub:categories} as well as \autoref{sub:language distribution}.

As we found many websites displaying alert boxes only to users with mobile web browser user agents and a relatively low diversity of messages (many of which are presented to one particular user agent only), our findings indicate that there are relatively few operators deploying their resources on a large number of different websites at the same time.

The distribution of unique messages and sites appears quite similar in the Austrian, US and Japanese scans. The notably higher number of distinct sites showing alert boxes only to devices with the user agent \texttt{iossafari} in the Austrian scan is caused by an accordingly increased count of sites displaying messages of the \textsc{Lottery} category, which is discussed in \autoref{sub:category analysis}.

\subsection{Category Analysis}
\label{sub:category analysis}

To make further analysis of the message content possible as well as to simplify the identification of specific phishing campaigns, we translated every message into English utilising Google Translate. This allowed us to classify the messages into the content categories described in \autoref{sub:categories} regardless of the original language. 

\begin{figure}[hbt!]
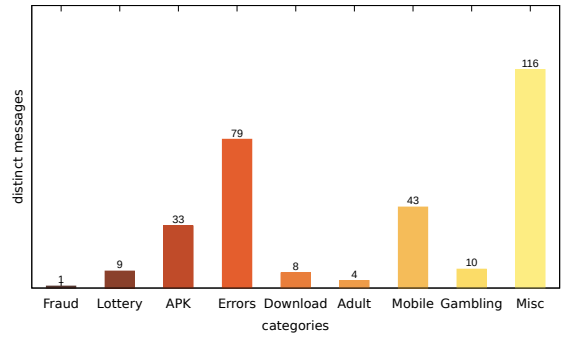

    \centering
     \begin{gnuplot}[terminal=pdf,scale=0.6]
            set boxwidth 0.5
            set style fill solid
            unset ytics
            set yrange [0:150]
            unset key
            set xlabel "categories"
            set ylabel "distinct messages"
            set linetype 1 lc rgb '#54362f'
            set linetype 2 lc rgb '#8a402b'
            set linetype 3 lc rgb '#c04b29'
            set linetype 4 lc rgb '#e45e2d'
            set linetype 5 lc rgb '#e97e3b'
            set linetype 6 lc rgb '#ef9c49'
            set linetype 7 lc rgb '#f5bc59'
            set linetype 8 lc rgb '#fbdc68'
            set linetype 9 lc rgb '#fded82'
            plot "Data/messages_per_category" using 1:3:($0+1):xtic(2) with boxes lc variable,"Data/messages_per_category" using 1:($3+3.6):3 with labels font "Arial,10"
        \end{gnuplot}
        \caption{Number of distinct messages per category (Austrian IP address)}
        \label{fig:messages_category}
\end{figure}

\begin{figure}[hbt!]
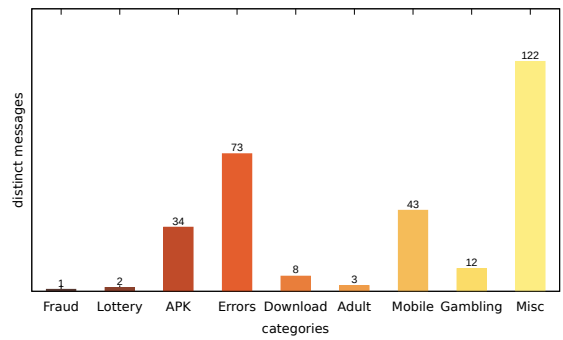

    \centering
     \begin{gnuplot}[terminal=pdf,scale=0.6]
            set boxwidth 0.5
            set style fill solid
            unset ytics
            set yrange [0:150]
            unset key
            set xlabel "categories"
            set ylabel "distinct messages"
            set linetype 1 lc rgb '#54362f'
            set linetype 2 lc rgb '#8a402b'
            set linetype 3 lc rgb '#c04b29'
            set linetype 4 lc rgb '#e45e2d'
            set linetype 5 lc rgb '#e97e3b'
            set linetype 6 lc rgb '#ef9c49'
            set linetype 7 lc rgb '#f5bc59'
            set linetype 8 lc rgb '#fbdc68'
            set linetype 9 lc rgb '#fded82'
            plot "Data_us/messages_per_category" using 1:3:($0+1):xtic(2) with boxes lc variable,"Data_us/messages_per_category" using 1:($3+3.6):3 with labels font "Arial,10"
        \end{gnuplot}
        \caption{Number of distinct messages per category (US IP address)}
        \label{fig:messages_category_us}
\end{figure}

\begin{figure}[hbt!]
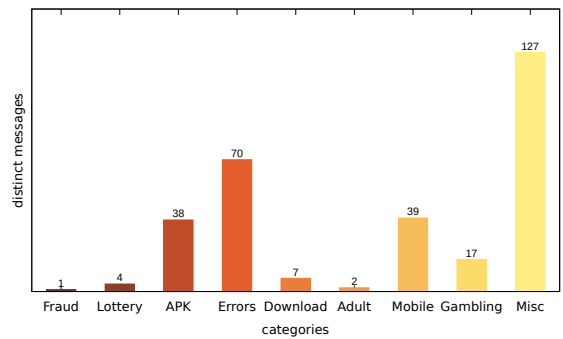

    \centering
    \begin{gnuplot}[terminal=pdf,scale=0.6]
            set boxwidth 0.5
            set style fill solid
            unset ytics
            set yrange [0:150]
            unset key
            set xlabel "categories"
            set ylabel "distinct messages"
            set linetype 1 lc rgb '#54362f'
            set linetype 2 lc rgb '#8a402b'
            set linetype 3 lc rgb '#c04b29'
            set linetype 4 lc rgb '#e45e2d'
            set linetype 5 lc rgb '#e97e3b'
            set linetype 6 lc rgb '#ef9c49'
            set linetype 7 lc rgb '#f5bc59'
            set linetype 8 lc rgb '#fbdc68'
            set linetype 9 lc rgb '#fded82'
            plot "Data_jp/messages_per_category" using 1:3:($0+1):xtic(2) with boxes lc variable,"Data_jp/messages_per_category" using 1:($3+3.6):3 with labels font "Arial,10"
        \end{gnuplot}
        \caption{Number of distinct messages per category (Japanese IP address)}
        \label{fig:messages_category_jp}
\end{figure}

Figures~\ref{fig:messages_category}, \ref{fig:messages_category_us} and \ref{fig:messages_category_jp} show the number of distinct messages in each category; as the figure shows, the greatest diversity in message content occurs in the categories \textsc{Misc}, \textsc{Mobile} and \textsc{Errors}.

\begin{figure}[hbt!]
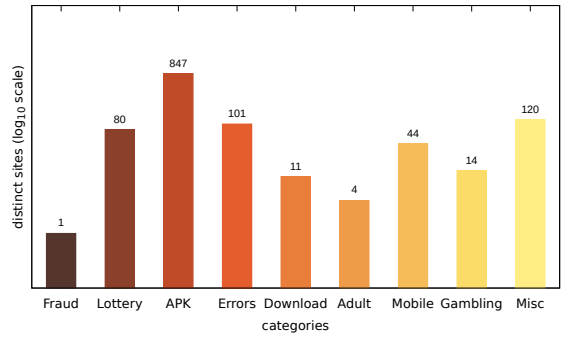

    \centering
   \begin{gnuplot}[terminal=pdf,scale=0.6]
            set boxwidth 0.5
            set style fill solid
            unset ytics
            set yrange [0.1:15000]
            unset key
            set logscale y
            set xlabel "categories"
            set ylabel "distinct sites (log_{10} scale)"
            set linetype 1 lc rgb '#54362f'
            set linetype 2 lc rgb '#8a402b'
            set linetype 3 lc rgb '#c04b29'
            set linetype 4 lc rgb '#e45e2d'
            set linetype 5 lc rgb '#e97e3b'
            set linetype 6 lc rgb '#ef9c49'
            set linetype 7 lc rgb '#f5bc59'
            set linetype 8 lc rgb '#fbdc68'
            set linetype 9 lc rgb '#fded82'
            plot "Data/sites_per_category" using 1:3:($0+1):xtic(2) with boxes lc variable,"Data/sites_per_category" using 1:($3+$3*0.6):3 with labels font "Arial,10"
        \end{gnuplot}
        \caption{Number of distinct sites per category (Austrian IP address)}
        \label{fig:sites_category}
\end{figure}

\begin{figure}[hbt!]
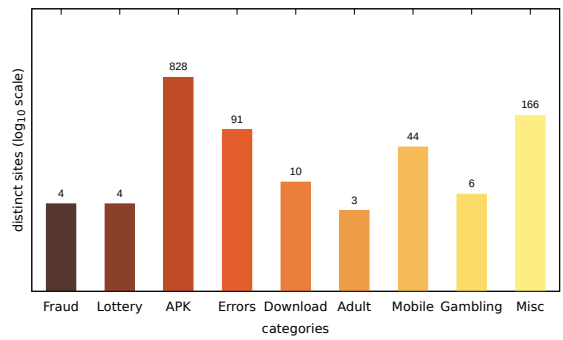

    \centering
    \begin{gnuplot}[terminal=pdf,scale=0.6]
            set boxwidth 0.5
            set style fill solid
            unset ytics
            set yrange [0.1:15000]
            unset key
            set logscale y
            set xlabel "categories"
            set ylabel "distinct sites (log_{10} scale)"
            set linetype 1 lc rgb '#54362f'
            set linetype 2 lc rgb '#8a402b'
            set linetype 3 lc rgb '#c04b29'
            set linetype 4 lc rgb '#e45e2d'
            set linetype 5 lc rgb '#e97e3b'
            set linetype 6 lc rgb '#ef9c49'
            set linetype 7 lc rgb '#f5bc59'
            set linetype 8 lc rgb '#fbdc68'
            set linetype 9 lc rgb '#fded82'
            plot "Data_us/sites_per_category" using 1:3:($0+1):xtic(2) with boxes lc variable,"Data_us/sites_per_category" using 1:($3+$3*0.6):3 with labels font "Arial,10"
        \end{gnuplot}
        \caption{Number of distinct sites per category (US IP address)}
        \label{fig:sites_category_us}
\end{figure}

\begin{figure}[hbt!]
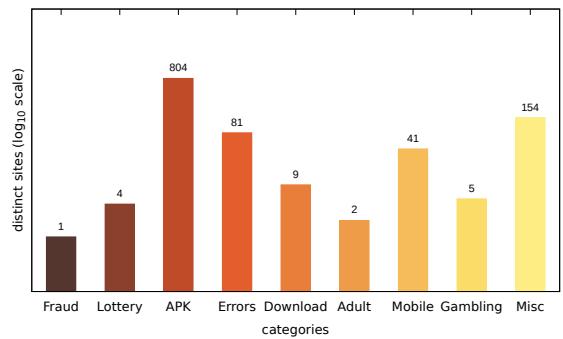

    \centering
    \begin{gnuplot}[terminal=pdf,scale=0.6]
            set boxwidth 0.5
            set style fill solid
            unset ytics
            set yrange [0.1:15000]
            unset key
            set logscale y
            set xlabel "categories"
            set ylabel "distinct sites (log_{10} scale)"
            set linetype 1 lc rgb '#54362f'
            set linetype 2 lc rgb '#8a402b'
            set linetype 3 lc rgb '#c04b29'
            set linetype 4 lc rgb '#e45e2d'
            set linetype 5 lc rgb '#e97e3b'
            set linetype 6 lc rgb '#ef9c49'
            set linetype 7 lc rgb '#f5bc59'
            set linetype 8 lc rgb '#fbdc68'
            set linetype 9 lc rgb '#fded82'
            plot "Data_jp/sites_per_category" using 1:3:($0+1):xtic(2) with boxes lc variable,"Data_jp/sites_per_category" using 1:($3+$3*0.6):3 with labels font "Arial,10"
        \end{gnuplot}
        \caption{Number of distinct sites per category (Japanese IP address)}
        \label{fig:sites_category_jp}
\end{figure}

While the majority of distinct messages belongs to legitimate non-malicious categories, most of the recorded alert boxes actually do fall into malicious categories. Figures~\ref{fig:sites_category}, \ref{fig:sites_category_us} and \ref{fig:sites_category_jp} depict the number of sites in each category. The vast majority of alert boxes belong to the category \textsc{APK} and tries to trick users into downloading and installing smartphone apps outside of the controlled environment of their OS's application store. Several ``distinct'' messages in the \textsc{Lottery} category were identical except for the current date being included inside the message text, which differs for the subsequent scans utilising the chosen user agents for technical reasons (i.\,e.\ the fact that we had to stagger our scans over several days). As a result, the number of distinct messages in this category appears somewhat higher in our data than it should be. Taking this into account would yield $4$ instead of $9$ distinct messages in the \textsc{Lottery} category for the Austrian scan and $1$ instead of $4$ distinct messages for the Japanese scan, while the count would remain the same for the US scan. The large number of distinct messages and websites in the \textsc{Misc} category is attributable to the scattered characteristics of the messages in this category and therefore not as significant.

Furthermore, we found noticeable differences between our previous scan and our new results. While the number of distinct messages in the \textsc{APK} category increased, the number of distinct sites displaying \textsc{APK} or \textsc{Lottery} messages reduced drastically. Distinct messages and sites in the \textsc{Fraud} category decreased as well. While most messages of the previous scan in this category claimed to provide some free credit for gambling if the user registers on the website, current messages try to either trick users into downloading and installing software or perform a technical support scam. 

Finally, we want to discuss the joint distribution of user agents and message content. Figures~\ref{fig:messages_category_useragent}, \ref{fig:messages_category_useragent_us} and \ref{fig:messages_category_useragent_jp} display the number of distinct messages shown only to a particular user agent for each of the malicious categories (\textsc{Fraud}, \textsc{Lottery}, and \textsc{APK}), whereas Figures~\ref{fig:sites_category_useragent}, \ref{fig:sites_category_useragent_us} and \ref{fig:sites_category_useragent_jp} show the corresponding graph for distinct websites. Most malicious alert boxes were encountered while utilising a mobile web browser user agent, while only a small fraction of the websites showed alert boxes on desktop browsers.

Our findings for the US and Japanese scans indicate that the majority of websites in the category \textsc{Lottery} targeted the user agent \texttt{iossafari}, while the Austrian scan showed a focus on \texttt{androidchrome} and \texttt{iossafari} (in that order). The websites in the category \textsc{APK} mostly targeted user agents of mobile web browsers (for obvious reasons), although there are few exceptions. Some websites do not distinguish between user agents and provide the APK file to all web browsers, while some seem to erroneously serve the APK file to a specific desktop user agent.


\subsection{Language Distribution}
\label{sub:language distribution}

Since we discovered a number of different trends for targeting specific user groups, we additionally analysed the language distribution of the collected messages. As shown in Figures~\ref{fig:languages_sites}, \ref{fig:languages_sites_us} and \ref{fig:languages_sites_jp}, the vast majority of websites displayed messages written in Chinese (zh) followed by English (en). The Austrian scan results show a significantly higher number of German (de) messages, while the increase in English messages is not as striking for the US scan. The Japanese scan results do not show a significant increase in websites displaying a Japanese (ja) message.

The main reason for the large number of Chinese messages is evident in Figures~\ref{fig:languages_categories}, \ref{fig:languages_categories_us}, \ref{fig:languages_categories_jp}, \ref{fig:languages_sites_categories}, \ref{fig:languages_sites_categories_us} and \ref{fig:languages_sites_categories_jp}, which show the distribution of languages over messages as well as websites in the different categories: The category \textsc{APK} (which contains the largest number of websites, cf.\ Figures~\ref{fig:sites_category}, \ref{fig:sites_category_us} and \ref{fig:sites_category_jp}) consists solely of messages in Chinese, and several other categories have a relatively large fraction of messages in Chinese, as well.

As we mentioned in our previous analysis, a large number of websites showing messages in the main language(s) of a country might indicate localisation attempts due to the utilisation of detection of IP addresses assigned to that country. The aforementioned figures provide an insight into the IP address-based localisation attempts, as well. Since most distinct German messages and according distinct sites contained in the Austrian scan results are categorised as \textsc{Lottery}, unlike the US and Japanese scan results, these seem to be localised for malicious purposes. In contrast, while the US scan does exhibit an increased number of distinct websites showing messages written in English, most of these are in the \textsc{Misc} category. The larger number of English messages in this scan hence does not provide significant evidence for localisation attempts, both due to the characteristics of the messages in this category and the distance of time between the Austrian and the US scans of each website (during which some websites could have added or changed messages, in particular informational messages falling into the \textsc{Misc} category). Finally, we were unable to detect any attempts at localisation in the Japanese scan results. Therefore our findings indicate that IP address-based targeting is only performed for specific languages.

The distribution of the languages across the different user agents is depicted in Figures~\ref{fig:language_messages_useragent}, \ref{fig:language_messages_useragent_us} and \ref{fig:language_messages_useragent_jp} as well as Figures~\ref{fig:language_sites_useragent}, \ref{fig:language_sites_useragent_us} and \ref{fig:language_sites_useragent_jp}.
While the largest fraction of distinct messages are written in English, the majority of websites display alert boxes with Chinese messages.

In summary, our results show that most alert boxes are displayed to visitors utilising web browsers with mobile user agents as well as that the majority of those alert box messages are in Chinese or English, with some country-specific evidence of localisation attempts based on users' IP addresses.

\section{Future Work}
\label{future work}

Compared to our previous study~\cite{popup_scam}, we explored language- and location-specific targeting by conducting three different scans utilising IP addresses from Austria, the United States and Japan via a VPN Service. Our results show that localisation is indeed applied by scammers, although apparently only for specific languages.

Additionally, we generated typosquatting domains based on the Majestic Million websites utilising \texttt{dnsmorph}, which is able to perform IDN homograph attacks -- which is an improvement over \texttt{dnstwist} used in our first scan. While the tool provided this functionality, our generated list did not contain any registered domains with IDN homographs.

A possibility to further improve our work could be the replacement of the manual process of assigning categories to distinct messages by a fully automated classification process utilising machine learning algorithms. A review of the existing categories might be necessary, as well, and the categorisation might rely solely on the message content instead of including background knowledge. This new process could establish a periodical automatic analysis enabling the observance of developments and trends in pop-up scams.

\section{Conclusion}
\label{sec:conclusion}

Techniques similar to those used for displaying pop-up ads in the early days of the World Wide Web are now used by malicious websites to deliver online scam. JavaScript alert message boxes steal the focus of the website, show a short text message to the user and try to either lure or scare the user into taking specific actions or exposing their data. Unfortunately, little scientific attention has been paid so far to the techniques utilised by scam websites to gain the attention of users and to retrieve data such as credit card information.

We performed large-scale scans of typosquatting URLs based on the Majestic Million websites with IP addresses from Austria, the United States and Japan via automated Chromium browsers utilising a modified version of the 
\textsc{MiningHunter} \cite{mininghunter} 
framework in combination with a VPN service. The three scans with five different user agents resulted in an aggregated total of $7\,400$ recorded alert boxes, out of which $4\,334$ can be considered malicious.

Our in-depth analysis presented characteristics of web-based scam campaigns and outlined target groups and goals of the various attacks. It showed that a significant fraction of websites displayed a pop-up box to one specific HTTP user agent only, and that most of them focused on mobile web browsers.

Different message categories were defined based on the message content and the websites displaying an alert box containing the message. The largest category is \textsc{APK}, which is trying to trick the user into directly downloading and installing a potentially malicious application.

Another aspect of our analysis was the distribution of different languages. We found that most websites were displaying alert box messages in Chinese, followed by English. Chinese messages often fell into the category \textsc{APK} and targeted a mobile web browser user agent.

Compared to our previous work, we found an overall decline of websites utilising malicious alert boxes for pop-up scam. The results of our Austrian scan showed the same evidence of localisation attempts, whereas the US and Japanese scans exhibited no significant evidence of localisation. It appears that this technique is only applied to some specific languages.

\section*{Acknowledgements}
This research was funded by the Austrian Research Promotion Agency (FFG) BRIDGE project 853264 ``Privacy and Security in Online Advertisement (PriSAd)'' as well as the Josef Ressel Center TARGET. The financial support by the Austrian Research Promotion Agency, the Federal Ministry for Digital and Economic Affairs and the Christian Doppler Research Association is gratefully acknowledged.

\appendix
\label{appendix}
\vspace{-1.0em}
\setlength{\belowcaptionskip}{-2.0em}
\captionsetup{justification=centering}

\begin{figure}[hbt!]
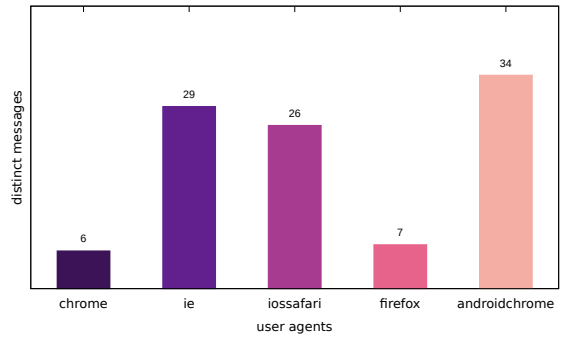

\centering
        \begin{gnuplot}[terminal=pdf,scale=0.6]
            set boxwidth 0.5
            set style fill solid
            unset ytics
            set yrange [0:45]
            unset key
            set xlabel "user agents"
            set ylabel "distinct messages"
            set linetype 1 lc rgb '#3c1357'
            set linetype 2 lc rgb '#61208d'
            set linetype 3 lc rgb '#a73b8f'
            set linetype 4 lc rgb '#e8638b'
            set linetype 5 lc rgb '#f4aea3'
            plot "Data/unique_messages_per_useragent" using 1:3:($0+1):xtic(2) title "unique urls per useragent" with boxes lc variable,"Data/unique_messages_per_useragent" using 1:($3+2):3 with labels font "Arial,10"
        \end{gnuplot}
        \caption{Number of distinct messages displayed to one \\ particular user agent (Austrian IP address)}
        \label{fig:uniquemessages}
\end{figure}
\begin{figure}[hbt!]
\centering
\begin{gnuplot}[terminal=pdf,scale=0.6]
            set boxwidth 0.5
            set style fill solid
            unset ytics
            set yrange [0:45]
            unset key
            set xlabel "user agents"
            set ylabel "distinct messages"
            set linetype 1 lc rgb '#3c1357'
            set linetype 2 lc rgb '#61208d'
            set linetype 3 lc rgb '#a73b8f'
            set linetype 4 lc rgb '#e8638b'
            set linetype 5 lc rgb '#f4aea3'
            plot "Data_us/unique_messages_per_useragent" using 1:3:($0+1):xtic(2) title "unique urls per useragent" with boxes lc variable,"Data_us/unique_messages_per_useragent" using 1:($3+2):3 with labels font "Arial,10"
        \end{gnuplot}
        \caption{Number of distinct messages displayed to one \\ particular user agent (US IP address)}
        \label{fig:uniquemessages_us}
\end{figure}
\begin{figure}[hbt!]
\centering
  \begin{gnuplot}[terminal=pdf,scale=0.6]
            set boxwidth 0.5
            set style fill solid
            unset ytics
            set yrange [0:45]
            unset key
            set xlabel "user agents"
            set ylabel "distinct messages"
            set linetype 1 lc rgb '#3c1357'
            set linetype 2 lc rgb '#61208d'
            set linetype 3 lc rgb '#a73b8f'
            set linetype 4 lc rgb '#e8638b'
            set linetype 5 lc rgb '#f4aea3'
            plot "Data_jp/unique_messages_per_useragent" using 1:3:($0+1):xtic(2) title "unique urls per useragent" with boxes lc variable,"Data_jp/unique_messages_per_useragent" using 1:($3+2):3 with labels font "Arial,10"
        \end{gnuplot}
        \caption{Number of distinct messages displayed to one \\ particular user agent (Japanese IP address)}
        \label{fig:uniquemessages_jp}
\end{figure}

\begin{figure}[hbt!]
    \centering
     \begin{gnuplot}[terminal=pdf,scale=0.6]
            set boxwidth 0.5
            set style fill solid
            set yrange [0.1:2500]
            set logscale y
            unset ytics
            unset key
            set xlabel "user agents"
            set ylabel "distinct sites (log_{10} scale)"
            set linetype 1 lc rgb '#3c1357'
            set linetype 2 lc rgb '#61208d'
            set linetype 3 lc rgb '#a73b8f'
            set linetype 4 lc rgb '#e8638b'
            set linetype 5 lc rgb '#f4aea3'
            plot "Data/unique_urls_per_useragent" using 1:3:($0+1):xtic(2) title "unique urls per useragent" with boxes lc variable,"Data/unique_urls_per_useragent" using 1:($3+$3*0.6):3 with labels font "Arial,10"
        \end{gnuplot}
        \caption{Number of websites targeting one \\ particular user agent (Austrian IP address)}
        \label{fig:uniquesites}
\end{figure}

\begin{figure}[hbt!]
    \centering
     \begin{gnuplot}[terminal=pdf,scale=0.6]
            set boxwidth 0.5
            set style fill solid
            set yrange [0.1:2500]
            set logscale y
            unset ytics
            unset key
            set xlabel "user agents"
            set ylabel "distinct sites (log_{10} scale)"
            set linetype 1 lc rgb '#3c1357'
            set linetype 2 lc rgb '#61208d'
            set linetype 3 lc rgb '#a73b8f'
            set linetype 4 lc rgb '#e8638b'
            set linetype 5 lc rgb '#f4aea3'
            plot "Data_us/unique_urls_per_useragent" using 1:3:($0+1):xtic(2) title "unique urls per useragent" with boxes lc variable,"Data_us/unique_urls_per_useragent" using 1:($3+$3*0.6):3 with labels font "Arial,10"
        \end{gnuplot}
        \caption{Number of websites targeting one \\ particular user agent (US IP address)}
        \label{fig:uniquesites_us}
\end{figure}

\begin{figure}[hbt!]
    \centering
    \begin{gnuplot}[terminal=pdf,scale=0.6]
    
        set style data histogram
        set style histogram cluster gap 1
        set ytics ("5" 5, "10" 10, "20" 20, "25" 25)
        set style fill solid
        set auto x
        set grid ytics
        set key left box
        set yrange [0:30]
        set xlabel "categories"
        set ylabel "distinct messages"
        plot 'Data/messages_per_category_per_useragent' using 2:xtic(1) title col lc rgb '#3c1357', \
                '' using 3:xtic(1) title col lc rgb '#61208d', \
                '' using 4:xtic(1) title col lc rgb '#a73b8f', \
                '' using 5:xtic(1) title col lc rgb '#e8638b', \
                '' using 6:xtic(1) title col lc rgb '#f4aea3'
        \end{gnuplot}
        \caption{Distribution of distinct messages over \\ different user agents in malicious categories (Austrian IP address)}
     \label{fig:messages_category_useragent}
\end{figure}

\begin{figure}[hbt!]
    \centering
     \begin{gnuplot}[terminal=pdf,scale=0.6]
    
        set style data histogram
        set style histogram cluster gap 1
        set ytics ("5" 5, "10" 10, "20" 20, "25" 25)
        set style fill solid
        set auto x
        set grid ytics
        set key left box
        set yrange [0:30]
        set xlabel "categories"
        set ylabel "distinct messages"
        plot 'Data_us/messages_per_category_per_useragent' using 2:xtic(1) title col lc rgb '#3c1357', \
                '' using 3:xtic(1) title col lc rgb '#61208d', \
                '' using 4:xtic(1) title col lc rgb '#a73b8f', \
                '' using 5:xtic(1) title col lc rgb '#e8638b', \
                '' using 6:xtic(1) title col lc rgb '#f4aea3'
        \end{gnuplot}
        \caption{Distribution of distinct messages over \\ different user agents in malicious categories (US IP address)}
     \label{fig:messages_category_useragent_us}
\end{figure}

\begin{figure}[hbt!]
    \centering
    \begin{gnuplot}[terminal=pdf,scale=0.6]
    
        set style data histogram
        set style histogram cluster gap 1
        set ytics ("5" 5, "10" 10, "20" 20, "25" 25)
        set style fill solid
        set auto x
        set grid ytics
        set key left box
        set yrange [0:30]
        set xlabel "categories"
        set ylabel "distinct messages"
        plot 'Data_jp/messages_per_category_per_useragent' using 2:xtic(1) title col lc rgb '#3c1357', \
                '' using 3:xtic(1) title col lc rgb '#61208d', \
                '' using 4:xtic(1) title col lc rgb '#a73b8f', \
                '' using 5:xtic(1) title col lc rgb '#e8638b', \
                '' using 6:xtic(1) title col lc rgb '#f4aea3'
        \end{gnuplot}
        \caption{Distribution of distinct messages over \\ different user agents in malicious categories (Japanese IP address)}
     \label{fig:messages_category_useragent_jp}
\end{figure}

\begin{figure}[hbt!]
    \centering
    \begin{gnuplot}[terminal=pdf,scale=0.6]
    
        set style data histogram
        set style histogram cluster gap 1
        set ytics ("5" 5 ,"10" 10, "50" 50, "100" 100, "500" 500, "800" 800)
        set style fill solid
        set auto x
        set grid ytics
        set logscale y
        set key left box
        set yrange [0.1:3000]
        set xlabel "categories"
        set ylabel "distinct sites (log_{10} scale)"
        plot 'Data/sites_per_category_per_useragent' using 2:xtic(1) title col lc rgb '#3c1357', \
                '' using 3:xtic(1) title col lc rgb '#61208d', \
                '' using 4:xtic(1) title col lc rgb '#a73b8f', \
                '' using 5:xtic(1) title col lc rgb '#e8638b', \
                '' using 6:xtic(1) title col lc rgb '#f4aea3'
        \end{gnuplot}
        \caption{Distribution of distinct sites over \\ different user agents in malicious categories (Austrian IP address)}
         \label{fig:sites_category_useragent}
\end{figure}

\begin{figure}[hbt!]
    \centering
   \begin{gnuplot}[terminal=pdf,scale=0.6]
    
        set style data histogram
        set style histogram cluster gap 1
        set ytics ("5" 5 ,"10" 10, "50" 50, "100" 100, "500" 500, "800" 800)
        set style fill solid
        set auto x
        set grid ytics
        set logscale y
        set key left box
        set yrange [0.1:3000]
        set xlabel "categories"
        set ylabel "distinct sites (log_{10} scale)"
        plot 'Data_us/sites_per_category_per_useragent' using 2:xtic(1) title col lc rgb '#3c1357', \
                '' using 3:xtic(1) title col lc rgb '#61208d', \
                '' using 4:xtic(1) title col lc rgb '#a73b8f', \
                '' using 5:xtic(1) title col lc rgb '#e8638b', \
                '' using 6:xtic(1) title col lc rgb '#f4aea3'
        \end{gnuplot}
        \caption{Distribution of distinct sites over \\ different user agents in malicious categories (US IP address)}
         \label{fig:sites_category_useragent_us}
\end{figure}

\begin{figure}[hbt!]
    \centering
   \begin{gnuplot}[terminal=pdf,scale=0.6]
    
        set style data histogram
        set style histogram cluster gap 1
        set ytics ("5" 5 ,"10" 10, "50" 50, "100" 100, "500" 500, "800" 800)
        set style fill solid
        set auto x
        set grid ytics
        set logscale y
        set key left box
        set yrange [0.1:3000]
        set xlabel "categories"
        set ylabel "distinct sites (log_{10} scale)"
        plot 'Data_jp/sites_per_category_per_useragent' using 2:xtic(1) title col lc rgb '#3c1357', \
                '' using 3:xtic(1) title col lc rgb '#61208d', \
                '' using 4:xtic(1) title col lc rgb '#a73b8f', \
                '' using 5:xtic(1) title col lc rgb '#e8638b', \
                '' using 6:xtic(1) title col lc rgb '#f4aea3'
        \end{gnuplot}
        \caption{Distribution of distinct sites over \\ different user agents in malicious categories (Japanese IP address)}
         \label{fig:sites_category_useragent_jp}
\end{figure}

\begin{figure}[hbt!]
    \centering
   \begin{gnuplot}[terminal=pdf,scale=0.6]
        set boxwidth 0.5
        set style fill solid
        unset ytics
        set yrange [0.1:3500]
        unset key
        set xlabel "languages (ISO 639-1)"
        set ylabel "distinct sites (log_{10} scale)"
        set logscale y
        set for [i=1:18] linetype i lc rgb '#d8d8d8'
        set linetype 3 lc rgb '#316ba7'
        set linetype 2 lc rgb '#223b89'
        set linetype 1 lc rgb '#151e5e'
        set linetype 5 lc rgb '#70c3d0'
        set linetype 4 lc rgb '#419dc5'
        plot "Data/languages_per_sites" using 1:3:($0+1):xtic(2) with boxes lc variable,"Data/languages_per_sites" using 1:($3+$3*0.6):3 with labels font "Arial,10"
    \end{gnuplot}
    \caption{Number of distinct sites displaying messages \\ in a specific language (Austrian IP address)}
    \label{fig:languages_sites}
\end{figure}

\begin{figure}[hbt!]
    \centering
   \begin{gnuplot}[terminal=pdf,scale=0.6]
        set boxwidth 0.5
        set style fill solid
        unset ytics
        set yrange [0.1:3500]
        unset key
        set xlabel "languages (ISO 639-1)"
        set ylabel "distinct sites (log_{10} scale)"
        set logscale y
        set for [i=1:18] linetype i lc rgb '#d8d8d8'
        set linetype 3 lc rgb '#316ba7'
        set linetype 2 lc rgb '#223b89'
        set linetype 1 lc rgb '#151e5e'
        set linetype 5 lc rgb '#70c3d0'
        set linetype 4 lc rgb '#419dc5'
        plot "Data_us/languages_per_sites" using 1:3:($0+1):xtic(2) with boxes lc variable,"Data_us/languages_per_sites" using 1:($3+$3*0.6):3 with labels font "Arial,10"
    \end{gnuplot}
    \caption{Number of distinct sites displaying messages \\ in a specific language (US IP address)}
    \label{fig:languages_sites_us}
\end{figure}

\begin{figure}[hbt!]
    \centering
   \begin{gnuplot}[terminal=pdf,scale=0.6]
        set boxwidth 0.5
        set style fill solid
        unset ytics
        set yrange [0.1:3500]
        unset key
        set xlabel "languages (ISO 639-1)"
        set ylabel "distinct sites (log_{10} scale)"
        set logscale y
        set for [i=1:18] linetype i lc rgb '#d8d8d8'
        set linetype 3 lc rgb '#316ba7'
        set linetype 2 lc rgb '#223b89'
        set linetype 1 lc rgb '#151e5e'
        set linetype 5 lc rgb '#70c3d0'
        set linetype 4 lc rgb '#419dc5'
        plot "Data_jp/languages_per_sites" using 1:3:($0+1):xtic(2) with boxes lc variable,"Data_jp/languages_per_sites" using 1:($3+$3*0.6):3 with labels font "Arial,10"
    \end{gnuplot}
    \caption{Number of distinct sites displaying messages \\ in a specific language (Japanese IP address)}
    \label{fig:languages_sites_jp}
\end{figure}

\begin{figure}[hbt!]
    \centering
    \begin{gnuplot}[terminal=pdf,scale=0.6]
        set style data histogram
        set style histogram cluster gap 1
        set ytics ("5" 5, "10" 10, "20" 20, "30" 30, "50" 50, "60" 60)
        set style fill solid
        set auto x
        set grid ytics
        set key left box
        set yrange [0:70]
        set xtics rotate by -45
        set xlabel "categories"
        set ylabel "distinct messages"
        plot 'Data/languages_per_unique_messages_per_all_category2' using 2:xtic(1) title col lc rgb '#151e5e', \
                '' using 3:xtic(1) title col lc rgb '#223b89', \
                '' using 4:xtic(1) title col lc rgb '#316ba7', \
                '' using 5:xtic(1) title col lc rgb '#d8d8d8'
        \end{gnuplot}
        \caption{Distribution of distinct messages over \\ different languages by category (Austrian IP address)}
        \label{fig:languages_categories}
\end{figure}

\begin{figure}[hbt!]
    \centering
    \begin{gnuplot}[terminal=pdf,scale=0.6]
        set style data histogram
        set style histogram cluster gap 1
        set ytics ("5" 5, "10" 10, "20" 20, "30" 30, "50" 50, "60" 60)
        set style fill solid
        set auto x
        set grid ytics
        set key left box
        set yrange [0:70]
        set xtics rotate by -45
        set xlabel "categories"
        set ylabel "distinct messages"
        plot 'Data_us/languages_per_unique_messages_per_all_category2' using 2:xtic(1) title col lc rgb '#151e5e', \
                '' using 3:xtic(1) title col lc rgb '#223b89', \
                '' using 4:xtic(1) title col lc rgb '#316ba7', \
                '' using 5:xtic(1) title col lc rgb '#d8d8d8'
        \end{gnuplot}
        \caption{Distribution of distinct messages over \\ different languages by category (US IP address)}
        \label{fig:languages_categories_us}
\end{figure}

\begin{figure}[hbt!]
    \centering
   \begin{gnuplot}[terminal=pdf,scale=0.6]
        set style data histogram
        set style histogram cluster gap 1
        set ytics ("5" 5, "10" 10, "20" 20, "30" 30, "50" 50, "60" 60)
        set style fill solid
        set auto x
        set grid ytics
        set key left box
        set yrange [0:70]
        set xtics rotate by -45
        set xlabel "categories"
        set ylabel "distinct messages"
        plot 'Data_jp/languages_per_unique_messages_per_all_category2' using 2:xtic(1) title col lc rgb '#151e5e', \
                '' using 3:xtic(1) title col lc rgb '#223b89', \
                '' using 4:xtic(1) title col lc rgb '#316ba7', \
                '' using 5:xtic(1) title col lc rgb '#d8d8d8'
        \end{gnuplot}
        \caption{Distribution of distinct messages over \\ different languages by category (Japanese IP address)}
        \label{fig:languages_categories_jp}
\end{figure}

\begin{figure}[hbt!]
    \centering
    \begin{gnuplot}[terminal=pdf,scale=0.6]
        set style data histogram
        set style histogram cluster gap 1
        set ytics ("5" 5 ,"10" 10, "50" 50, "100" 100, "800" 800)
        set style fill solid
        set auto x
        set grid ytics
        set key left box
        set yrange [0.1:3000]
        set xtics rotate by -45
        set xlabel "categories"
        set ylabel "distinct sites (log_{10} scale)"
        set logscale y
        plot 'Data/languages_per_sites_per_category' using 2:xtic(1) title col lc rgb '#151e5e', \
                '' using 3:xtic(1) title col lc rgb '#223b89', \
                '' using 4:xtic(1) title col lc rgb '#316ba7', \
                '' using 5:xtic(1) title col lc rgb '#d8d8d8'
        \end{gnuplot}
        \caption{Distribution of distinct sites over \\ different languages by category (Austrian IP address)}
        \label{fig:languages_sites_categories}
\end{figure}

\begin{figure}[hbt!]
    \centering
   \begin{gnuplot}[terminal=pdf,scale=0.6]
        set style data histogram
        set style histogram cluster gap 1
        set ytics ("5" 5 ,"10" 10, "50" 50, "100" 100, "800" 800)
        set style fill solid
        set auto x
        set grid ytics
        set key left box
        set yrange [0.1:3000]
        set xtics rotate by -45
        set xlabel "categories"
        set ylabel "distinct sites (log_{10} scale)"
        set logscale y
        plot 'Data_us/languages_per_sites_per_category' using 2:xtic(1) title col lc rgb '#151e5e', \
                '' using 3:xtic(1) title col lc rgb '#223b89', \
                '' using 4:xtic(1) title col lc rgb '#316ba7', \
                '' using 5:xtic(1) title col lc rgb '#d8d8d8'
        \end{gnuplot}
        \caption{Distribution of distinct sites over \\ different languages by category (US IP address)}
        \label{fig:languages_sites_categories_us}
\end{figure}

\begin{figure}[hbt!]
    \centering
   \begin{gnuplot}[terminal=pdf,scale=0.6]
        set style data histogram
        set style histogram cluster gap 1
        set ytics ("5" 5 ,"10" 10, "50" 50, "100" 100, "800" 800)
        set style fill solid
        set auto x
        set grid ytics
        set key left box
        set yrange [0.1:3000]
        set xtics rotate by -45
        set xlabel "categories"
        set ylabel "distinct sites (log_{10} scale)"
        set logscale y
        plot 'Data_jp/languages_per_sites_per_category' using 2:xtic(1) title col lc rgb '#151e5e', \
                '' using 3:xtic(1) title col lc rgb '#223b89', \
                '' using 4:xtic(1) title col lc rgb '#316ba7', \
                '' using 5:xtic(1) title col lc rgb '#d8d8d8'
        \end{gnuplot}
        \caption{Distribution of distinct sites over \\ different languages by category (Japanese IP address)}
        \label{fig:languages_sites_categories_jp}
\end{figure}

\begin{figure}[hbt!]
    \centering
       \begin{gnuplot}[terminal=pdf,scale=0.6]
        set style data histogram
        set style histogram cluster gap 1
        set ytics ("5" 5, "20" 20, "30" 30, "40" 40, "60" 60, "80" 80)
        set style fill solid
        set auto x
        set grid ytics
        set key left box
        set yrange [0:120]
        set xlabel "user agents"
        set ylabel "distinct messages"
        plot 'Data/languages_per_unique_messages_per_useragent2' using 2:xtic(1) title col lc rgb '#151e5e', \
                '' using 3:xtic(1) title col lc rgb '#223b89', \
                '' using 4:xtic(1) title col lc rgb '#316ba7', \
                '' using 5:xtic(1) title col lc rgb '#419dc5', \
                '' using 6:xtic(1) title col lc rgb '#70c3d0', \
                '' using 7:xtic(1) title col lc rgb '#d8d8d8'
        \end{gnuplot}
        \caption{Distribution of distinct messages over \\ different languages by user agent (Austrian IP address)}
        \label{fig:language_messages_useragent}
\end{figure}

\begin{figure}[hbt!]
    \centering
    \begin{gnuplot}[terminal=pdf,scale=0.6]
        set style data histogram
        set style histogram cluster gap 1
        set ytics ("5" 5, "20" 20, "30" 30, "40" 40, "60" 60, "80" 80)
        set style fill solid
        set auto x
        set grid ytics
        set key left box
        set yrange [0:120]
        set xlabel "user agents"
        set ylabel "distinct messages"
        plot 'Data_us/languages_per_unique_messages_per_useragent2' using 2:xtic(1) title col lc rgb '#151e5e', \
                '' using 3:xtic(1) title col lc rgb '#223b89', \
                '' using 4:xtic(1) title col lc rgb '#316ba7', \
                '' using 5:xtic(1) title col lc rgb '#419dc5', \
                '' using 6:xtic(1) title col lc rgb '#70c3d0', \
                '' using 7:xtic(1) title col lc rgb '#d8d8d8'
        \end{gnuplot}
        \caption{Distribution of distinct messages over \\ different languages by user agent (US IP address)}
        \label{fig:language_messages_useragent_us}
\end{figure}

\begin{figure}[hbt!]
    \centering
    \begin{gnuplot}[terminal=pdf,scale=0.6]
        set style data histogram
        set style histogram cluster gap 1
        set ytics ("5" 5, "20" 20, "30" 30, "40" 40, "60" 60, "80" 80)
        set style fill solid
        set auto x
        set grid ytics
        set key left box
        set yrange [0:120]
        set xlabel "user agents"
        set ylabel "distinct messages"
        plot 'Data_jp/languages_per_unique_messages_per_useragent2' using 2:xtic(1) title col lc rgb '#151e5e', \
                '' using 3:xtic(1) title col lc rgb '#223b89', \
                '' using 4:xtic(1) title col lc rgb '#316ba7', \
                '' using 5:xtic(1) title col lc rgb '#419dc5', \
                '' using 6:xtic(1) title col lc rgb '#70c3d0', \
                '' using 7:xtic(1) title col lc rgb '#d8d8d8'
        \end{gnuplot}
        \caption{Distribution of distinct messages over \\ different languages by user agent (Japanese IP address)}
        \label{fig:language_messages_useragent_jp}
\end{figure}

\begin{figure}[hbt!]
    \centering
       \begin{gnuplot}[terminal=pdf,scale=0.6]
        set style data histogram
        set style histogram cluster gap 1
         set ytics ("5" 5 ,"10" 10, "50" 50, "100" 100, "500" 500, "800" 800)
        set style fill solid
        set auto x
        set grid ytics
        set key left box
        set yrange [0.1:3000]
        set xlabel "user agents"
        set ylabel "distinct sites (log_{10} scale)"
        set logscale y
        plot 'Data/languages_per_sites_per_useragent' using 2:xtic(1) title col lc rgb '#151e5e', \
                '' using 3:xtic(1) title col lc rgb '#223b89', \
                '' using 4:xtic(1) title col lc rgb '#316ba7', \
                '' using 5:xtic(1) title col lc rgb '#419dc5', \
                '' using 6:xtic(1) title col lc rgb '#70c3d0', \
                '' using 7:xtic(1) title col lc rgb '#d8d8d8'
        \end{gnuplot}
        \caption{Distribution of distinct sites over \\ different languages by user agent (Austrian IP address)}
        \label{fig:language_sites_useragent}
\end{figure}

\begin{figure}[hbt!]
    \centering
     \begin{gnuplot}[terminal=pdf,scale=0.6]
        set style data histogram
        set style histogram cluster gap 1
         set ytics ("5" 5 ,"10" 10, "50" 50, "100" 100, "500" 500, "800" 800)
        set style fill solid
        set auto x
        set grid ytics
        set key left box
        set yrange [0.1:3000]
        set xlabel "user agents"
        set ylabel "distinct sites (log_{10} scale)"
        set logscale y
        plot 'Data_us/languages_per_sites_per_useragent' using 2:xtic(1) title col lc rgb '#151e5e', \
                '' using 3:xtic(1) title col lc rgb '#223b89', \
                '' using 4:xtic(1) title col lc rgb '#316ba7', \
                '' using 5:xtic(1) title col lc rgb '#419dc5', \
                '' using 6:xtic(1) title col lc rgb '#70c3d0', \
                '' using 7:xtic(1) title col lc rgb '#d8d8d8'
        \end{gnuplot}
        \caption{Distribution of distinct sites over \\ different languages by user agent (US IP address)}
        \label{fig:language_sites_useragent_us}
\end{figure}

\begin{figure}[hbt!]
    \centering
    \begin{gnuplot}[terminal=pdf,scale=0.6]
        set style data histogram
        set style histogram cluster gap 1
         set ytics ("5" 5 ,"10" 10, "50" 50, "100" 100, "500" 500, "800" 800)
        set style fill solid
        set auto x
        set grid ytics
        set key left box
        set yrange [0.1:3000]
        set xlabel "user agents"
        set ylabel "distinct sites (log_{10} scale)"
        set logscale y
        plot 'Data_jp/languages_per_sites_per_useragent' using 2:xtic(1) title col lc rgb '#151e5e', \
                '' using 3:xtic(1) title col lc rgb '#223b89', \
                '' using 4:xtic(1) title col lc rgb '#316ba7', \
                '' using 5:xtic(1) title col lc rgb '#419dc5', \
                '' using 6:xtic(1) title col lc rgb '#70c3d0', \
                '' using 7:xtic(1) title col lc rgb '#d8d8d8'
        \end{gnuplot}
        \caption{Distribution of distinct sites over \\ different languages by user agent (Japanese IP address)}
        \label{fig:language_sites_useragent_jp}
\end{figure}
\bibliographystyle{plain}
\bibliography{Bibliography}

\end{document}